\documentclass[12pt]{article}
\usepackage{hyperref}
\usepackage[english]{babel}
\usepackage{amsmath}
\usepackage{latexsym}
\usepackage{amssymb}
\usepackage[all]{xy}
\usepackage[dvips]{graphicx}
\usepackage{epsfig}
\usepackage{psfrag}

\numberwithin{equation}{section} \numberwithin{table}{section}
\numberwithin{figure}{section}

\begin{document}



\begin{titlepage}

  \begin{center}

    \vspace{20mm}

    {\LARGE \bf Holography of Charged Black Holes with $RF^2$ Corrections}

    \vspace{10mm}

   Rong-Gen Cai$^{\ast}$ and Da-Wei Pang$^{\sharp\ast}$

    \vspace{5mm}
    {\small \sl $\ast$ Key Laboratory of Frontiers in Theoretical Physics,\\
    Institute of Theoretical Physics, Chinese Academy of Sciences, \\
    P.O. Box 2735, Beijing 100190, China\\}
     {\small \sl $\sharp$ CENTRA, Departamento de F\'{\i}sica,}\\
    {\small \sl Instituto Superior T\'{e}cnico,
Universidade T\'{e}cnica de Lisboa}\\
    {\small \sl Av. Rovisco Pais 1, 1049-001 Lisboa, Portugal}\\
    {\small \tt cairg@itp.ac.cn, dwpang@gmail.com}
    \vspace{10mm}

  \end{center}

\begin{abstract}
\baselineskip=18pt We investigate several holographic properties of
charged black holes with $RF^{2}$ corrections, which originate from
Kaluza-Klein reductions of five-dimensional Gauss-Bonnet gravity. We
obtain the perturbative solution and discuss its thermodynamics. The
DC conductivity is calculated via the effective action approach and
the corrections to the DC conductivity are also evaluated. Moreover,
we calculate the shear viscosity $\eta$ and thermal conductivity
$\kappa_{T}$, as well as two interesting ratios $\eta/s$ and
$\kappa_{T}\mu^{2}/(\eta T)$. We find that both bounds conjectured
in the literatures can be violated in the presence of the higher-
order corrections $RF^2$.

\end{abstract}
\setcounter{page}{0}
\end{titlepage}

\pagestyle{plain} \baselineskip=19pt

\tableofcontents

\section{Introduction}
The AdS/CFT correspondence~\cite{Maldacena:1997re, Aharony:1999ti} has proven to be a
powerful tool for studying the dynamics of strongly coupled field theories. It is widely
recognized that this framework has opened a new window toward understanding real-world physics,
such as QCD. Moreover, in recent years investigations on applications of AdS/CFT correspondence
to condensed matter physics (or AdS/CMT for short) have accelerated enormously. For reviews, see~\cite{Hartnoll:2009sz}.

In order to describe condensed matter systems holographically, the
minimum set of fields in the dual gravity side contains the graviton
$g_{ab}$, the $U(1)$ gauge field $A_{a}$, the fermion $\psi$ and/or
the dilaton $\phi$. Thus according to the AdS/CFT dictionary, the
bulk configuration is dual to a certain system with finite charge
density, either at finite temperature or at extremality. The
simplest corresponding black hole solution is the
Reissner-Nordstr\"{o}m black hole in four-dimensional anti-de Sitter
spacetime (RN-${\rm AdS_{4}}$). It has been observed that the
RN-${\rm AdS}_{4}$ black hole is a perfect laboratory for
investigating this AdS/CMT correspondence. In particular, studies on
the fermionic correlation functions in this background revealed the
existence of fermionic quasi-particles with non-Fermi liquid
behavior ~\cite{Lee:2008xf, Liu:2009dm, Cubrovic:2009ye}. Moreover,
it was shown in~\cite{Faulkner:2009wj} that in the extremal RN-${\rm
AdS}_{4}$ background, the scaling behavior of certain correlation
functions was governed by an emergent IR CFT, which was associated
with the ${\rm AdS}_{2}$ symmetry in the near-horizon geometry.
Furthermore, such a system also exhibited superconducting phase
transitions upon the introduction of charged
scalar hair~\cite{Gubser:2008px, Hartnoll:2008vx, Hartnoll:2008kx}.

One particularly interesting quantity characterizing charge transport properties
of a certain condensed matter system is the conductivity. According to the AdS/CFT correspondence,
it can be evaluated in the dual gravity side by working out the current-current correlation
function of the bulk $U(1)$ gauge field. In~\cite{Herzog:2007ij} the authors considered
charge transport properties of 2+1 dimensional CFTs at non-zero temperature, where they found
that the electromagnetic self-duality of the 3+1 dimensional theory implied that
the the R-current correlation functions obeyed some constraint similar to
that found from particle-vortex duality in 2+1 dimensional Abelian theories.

One may ask if this self-duality still holds in the presence of higher-order corrections.
This issue was explored recently in~\cite{Myers:2010pk}, where the authors considered
a particular form of the higher-order corrections, i.e. the Weyl tensor coupled to gauge field strengths.
They calculated corrections to the conductivity and found that although self-duality was lost in
higher-derivative theories, a similar relation still held for the transverse and longitudinal
components of the charge correlation functions, which reduced to the self-dual version at leading order.

The form of the higher-order corrections is not unique, then one wonders how charge transport properties
will be modified if other forms of higher-order corrections are considered. In~\cite{Myers:2010pk} higher-order corrections with $RF^{2}$ terms were also discussed, whose details will be reviewed in the next section.
It was shown that in neutral black hole background, the two actions were equivalent up to redefinitions
of coupling constants, which led to the same charge transport properties.

A natural question concerning the charge transport properties is what will happen at finite charge density.
This is the main focus of this paper, i.e. to discuss holographic properties of charged black holes in the
presence of $RF^{2}$ corrections. In section 2 we first give a brief review on~\cite{Myers:2010pk}.
Then we obtain the corrected background via perturbative techniques and discuss its thermodynamic
properties in section 3. The DC conductivity is calculated in section 4, following the effective action approach presented in~\cite{Myers:2009ij}. We find that when the charge parameter $Q=0$, the DC conductivity agrees with the one discussed in~\cite{Myers:2010pk}. Next we consider the shear viscosity $\eta$ and thermal conductivity
$\kappa_{T}$ in section 5 and compute two interesting ratios $\eta/s$ and $\kappa_{T}\mu^{2}/(\eta T)$.
We find that both ratios can be modified by higher-order corrections, thus the corresponding bounds
discussed in previous literatures may be violated. Finally we will summarize our results and discuss
relevant issues in the last section.
\section{Preliminaries: charge transport in neutral black hole background}
It was found in~\cite{Herzog:2007ij} that the conductivity in three-dimensional field theory
side at zero momentum was a constant with no frequency dependence.
The authors of~\cite{Herzog:2007ij} attributed this remarkable result to the electromagnetic
self-duality of the four-dimensional bulk Einstein-Maxwell theory. Recently
in order to acquire a better understanding of this self-duality, the authors of~\cite{Myers:2010pk}
considered a particular form of new higher-derivative corrections which involved couplings between
the gauge field to the spacetime curvature. Part of the main results are summarized as follows.

The starting point was the four-dimensional planar Schwarzschild-${\rm AdS}_{4}$ black hole,
$$ds^{2}=\frac{r^{2}}{L^{2}}(-f(r)dt^{2}+dx^{2}+dy^{2})+\frac{L^{2}dr^{2}}{r^{2}f(r)},$$
where $f(r)=1-r^{3}_{0}/r^{3}$. On the other hand, after integrating by parts and imposing
the identities $\nabla_{[a}F_{bc]}=R_{[abc]d}=0$, the most general four-derivative action
contains the following terms,
\begin{eqnarray}
\label{equ1}
I_{4}&=&\int d^{4}x\sqrt{-g}[\alpha_{1}R^{2}+\alpha_{2} R_{ab}R^{ab}+\alpha_{3}(F^{2})^{2}+\alpha_{4}F^{4}
+\alpha_{5}\nabla^{a}F_{ab}\nabla^{c}{F_{c}}^{b}\nonumber\\
& &~~~~~~~~~~~~~+\alpha_{6}R_{abcd}F^{ab}F^{cd}+\alpha_{7}R^{ab}F_{ac}{F_{b}}^{c}+\alpha_{8}RF^{2}],
\end{eqnarray}
where $F^{2}=F_{ab}F^{ab}, F^{4}={F^{a}}_{b}{F^{b}}_{c}{F^{c}}_{d}{F^{d}}_{a}$.
If we focus on the conductivity, which means that only the current-current two-point functions
are relevant, we can just consider the effects of the $\alpha_{6}, \alpha_{7}$ and $\alpha_{8}$
terms. Furthermore, after taking a particular linear combination of these three terms, the effective
action for bulk Maxwell field turns out to be
\begin{equation}
\label{equ2}
I_{\rm vec}=\frac{1}{g^{2}_{4}}\int d^{4}x\sqrt{-g}[-\frac{1}{4}F_{ab}F^{ab}+\gamma L^{2}C_{abcd}F^{ab}F^{cd}],
\end{equation}
where $C_{abcd}$ denotes the Weyl tensor.
Then the DC conductivity in the presence of higher-order corrections is given by
\begin{equation}
\sigma_{\rm DC}=\frac{1}{g^{2}_{4}}(1+4\gamma).
\end{equation}

One can also consider the following type of higher-order corrections instead
\begin{equation}
\label{equ4}
I^{\prime}_{\rm vec}=\frac{1}{\tilde{g}^{2}_{4}}\int d^{4}x\sqrt{-g}[-\frac{1}{4}F_{ab}F^{ab}+
\alpha L^{2}(R_{abcd}F^{ab}F^{cd}-4R_{ab}F^{ac}{F^{b}}_{c}+RF^{ab}F_{ab})],
\end{equation}
which arises from the Kaluza-Klein reduction of five-dimensional Gauss-Bonnet gravity.
It was observed in~\cite{Myers:2010pk} that by combining the Einstein equation in
the neutral black hole background $R_{ab}=-3/L^{2}g_{ab}$ and the definition
of the Weyl tensor, the action~(\ref{equ4}) becomes
\begin{equation}
I^{\prime}_{\rm vec}=\frac{1+8\alpha}{\tilde{g}^{2}_{4}}\int d^{4}x\sqrt{-g}[-\frac{1}{4}F_{ab}F^{ab}
+\frac{\alpha}{1+8\alpha}L^{2}C_{abcd}F^{ab}F^{cd}].
\end{equation}
It can be easily seen that the resulting action is equivalent to~(\ref{equ2}) with the following identifications
\begin{equation}
g^{2}_{4}=\frac{\tilde{g}^{2}_{4}}{1+8\alpha},~~~\gamma=\frac{\alpha}{1+8\alpha}.
\end{equation}
Therefore the charge transport properties are identical. In particular, the
DC conductivity can be re-expressed as
\begin{equation}
\sigma_{\rm DC}=\frac{1+12\alpha}{\tilde{g}^{2}_{4}}.
\end{equation}

Moreover, gravity duals at finite charge density are of particular interest especially in the context
of AdS/CMT. Then it would be desirable to study the effects of such higher-order corrections. However,
due to the presence of the nontrivial $U(1)$ background gauge field, (\ref{equ2}) and~(\ref{equ4})
are no longer equivalent. Then to consider higher-order corrections to the conductivity, one has to
take the back-reaction of the $U(1)$ gauge field into account. In the next section we will calculate
the perturbative solution with $RF^{2}$ corrections and discuss its thermodynamics.
\section{The perturbative solution with $RF^{2}$ corrections}
Our staring point is the RN-${\rm AdS}_{4}$ black hole with planar symmetry, whose action is given by
\begin{equation}
S_{0}=\frac{1}{2\kappa^{2}}\int d^{4}x\sqrt{-g}[R+\frac{6}{L^{2}}-\frac{L^{2}}{g^{2}_{F}}F_{ab}F^{ab}].
\end{equation}
The black hole metric can be expressed as
\begin{equation}
ds_{0}^{2}=\frac{r^{2}}{L^{2}}[-f_{0}(r)dt^{2}+dx^{2}+dy^{2}]+\frac{L^{2}}{r^{2}}\frac{dr^{2}}{f_{0}(r)},
\end{equation}
where
\begin{equation}
f_{0}(r)=1-\frac{M}{r^{3}}+\frac{Q^{2}}{r^{4}}.
\end{equation}
There exists a nontrivial background gauge field
\begin{equation}
A^{(0)}_{t}=\mu_{0}(1-\frac{r_{0}}{r}).
\end{equation}
Note that we put a subscript `0' to denote the leading order quantities from now on.
The horizon locates at $r=r_{0}$, which satisfies $f_{0}(r_{0})=0$. Then we have $M=r_{0}^{3}+Q^{2}/r_{0}$.
The thermodynamic quantities, such as the chemical potential $\mu_{0}$, the charge density $\rho_{0}$, the energy density $\epsilon_{0}$ and the entropy density $s_{0}$, are given as follows
\begin{eqnarray}
& &\mu_{0}=\frac{g_{F}Q}{L^{2}r_{0}},~~~\rho_{0}=\frac{2Q}{\kappa^{2}L^{2}g_{F}},\nonumber\\
& &\epsilon_{0}=\frac{M}{\kappa^{2}L^{4}},~~~s_{0}=\frac{2\pi r_{0}^{2}}{\kappa^{2}L^{2}}.
\end{eqnarray}
We can also evaluate the temperature
\begin{equation}
T_{0}=\frac{3r_{0}}{4\pi L^{2}}(1-\frac{Q^{2}}{3r_{0}^{4}}),
\end{equation}
When we take the extremal limit we can obtain
\begin{equation}
T_{0}=0,~~~\Rightarrow~~~Q^{2}=3r^{4}_{0}.
\end{equation}
It can be easily verified that the first law of thermodynamics holds
\begin{equation}
d\epsilon_{0}=T_{0}ds_{0}+\mu_{0}d\rho_{0}.
\end{equation}

Next we consider the effects of the higher-order corrections. As illustrated in the introduction,
due to the presence of the background $U(1)$ gauge field, one should work out the back-reacted solution at first.
However, generically it is always difficult to obtain exact solutions in theories with higher-order corrections.
Then we may try to obtain the perturbative solution, following~\cite{Myers:2009ij}. Black holes in five-dimensional
gauged supergravity with higher-derivatives were also studied in~\cite{Cremonini:2008tw}. Here the full action
including higher-order corrections is given as follows
\begin{eqnarray}
S&\equiv&S_{0}+S_{1}\nonumber\\
&=&\frac{1}{2\kappa^{2}}\int d^{4}x\sqrt{-g}[R+\frac{6}{L^{2}}-\frac{L^{2}}{g^{2}_{F}}F_{ab}F^{ab}\nonumber\\
& &+\frac{\alpha L^{4}}{g^{2}_{F}}(R_{abcd}F^{ab}F^{cd}-4R_{ab}F^{ac}{F^{b}}_{c}+RF^{ab}F_{ab})],
\end{eqnarray}
where we have rewritten the form of the correction terms so that
$\alpha$ is a dimensionless constant. It would be straightforward to
obtain the modified Maxwell equation
\begin{equation}
\nabla_{a}[F^{ab}-\alpha L^{2}({R^{ab}}_{cd}F^{cd}-2R^{ac}{F_{c}}^{b}+2R^{bc}{F_{c}}^{a}+RF^{ab})]=0
\end{equation}
as well as the Einstein equation
\begin{equation}
R_{ab}-\frac{1}{2}Rg_{ab}=\frac{3}{L^{2}}+\frac{2L^{2}}{g^{2}_{F}}F_{ac}{F_{b}}^{c}
-\frac{L^{2}}{2g^{2}_{F}}g_{ab}F^{2}
+\frac{\alpha L^{4}}{g^{2}_{F}}(G_{1ab}+G_{2ab}+G_{3ab}),
\end{equation}
where
\begin{eqnarray}
G_{1ab}&=&\frac{1}{2}g_{ab}R_{cdef}F^{cd}F^{ef}-2R_{adef}{F_{b}}^{d}F^{ef}
-2R_{bdef}{F_{a}}^{d}F^{ef}+2\nabla^{d}\nabla^{f}F_{da}F_{bf},\nonumber\\
G_{2ab}&=&-2g_{ab}R_{cd}F^{ce}{F^{d}}_{e}-2\nabla_{d}\nabla_{a}F_{bf}F^{df}
-2\nabla_{d}\nabla_{b}F_{af}F^{df}\nonumber\\
& &+2\Box F_{af}{F_{b}}^{f}+2g_{ab}\nabla_{c}\nabla_{d}{F^{c}}_{f}F^{df}\nonumber\\
& &+4R_{ac}F_{bf}F^{cf}+4R_{bc}F_{af}F^{cf}+2R_{cd}{F^{c}}_{a}{F^{d}}_{b}+2R_{cd}{F^{c}}_{b}{F^{d}}_{a},\nonumber\\
G_{3ab}&=&\frac{1}{2}g_{ab}RF^{2}-R_{ab}F^{2}+\nabla_{a}\nabla_{b}F^{2}
+g_{ab}\Box F^{2}+2RF_{ac}{F_{b}}^{c}.
\end{eqnarray}

We take the following ansatz for the perturbative solution
\begin{equation}
ds^{2}=\frac{r^{2}}{L^{2}}[-f(r)dt^{2}+dx^{2}+dy^{2}]+\frac{L^{2}}{r^{2}}\frac{dr^{2}}{g(r)},~~~
A_{t}(r)=A^{(0)}_{t}(r)+H(r),
\end{equation}
where
\begin{equation}
f(r)=f_{0}(r)(1+F(r)),~~~g(r)=f_{0}(r)(1+F(r)+G(r)).
\end{equation}
The main steps proposed in~\cite{Myers:2009ij} for calculating the perturbative solution can be
summarized as follows:
\begin{enumerate}
\item Considering the combination $G^{t}_{t}-f/gG^{r}_{r}$, where $G_{ab}$ denotes the Einstein tensor,
one finds a first-order linear ODE for $G(r)$, which is solvable.
\item Given $G(r)$, the modified Maxwell equation is easily solved.
\item With the above two perturbations, $F(r)$ can be determined by solving the first-order linear
ODE coming from the $rr$ component of the Einstein equation.
\end{enumerate}
Let us turn to our particular ansatz. It can be seen that step 1 gives
\begin{equation}
rf_{0}(r)\partial_{r}G(r)=0,~~~\Rightarrow~~~\partial_{r}G(r)=0,~~G(r)={\rm const}.
\end{equation}
Without loss of generality we can set $G(r)=0$, therefore
\begin{equation}
f(r)=g(r)=f_{0}(r)(1+F(r)).
\end{equation}
Subsequently we can obtain the perturbation for the gauge field from step 2
\begin{equation}
H(r)=h_{0}+\frac{h_{1}}{r}+\frac{2\alpha\mu_{0}r_{0}}{r}-\frac{\alpha\mu_{0}r^{4}_{0}}{2r^{4}}-
\frac{\alpha\mu_{0}Q^{2}}{2r^{4}}+\frac{2\alpha\mu_{0}r_{0}Q^{2}}{5r^{5}}.
\end{equation}
Finally by combining the above two functions, the perturbation for the metric component is
given by
\begin{equation}
Y(r)\equiv f_{0}(r)F(r)=\frac{y_{0}}{r^{3}}-\frac{\alpha Q^{2}r_{0}^{3}}{2r^{7}}
-\frac{\alpha Q^{4}}{2r_{0}r^{7}}+\frac{8\alpha Q^{4}}{5r^{8}}.
\end{equation}

To determine the integration constants in the above perturbative
solutions, we can impose several constraints
following~\cite{Myers:2009ij}:
\begin{itemize}
\item First we require that $r=r_{0}$ is still the horizon, which leads to $Y(r_{0})=0$.
Then the constant $y_{0}$ is given by
\begin{equation}
y_{0}=\frac{\alpha Q^{2}}{2r_{0}}-\frac{11\alpha Q^{4}}{10r_{0}^{5}}.
\end{equation}
\item The second requirement is that the gauge potential $A_{t}(r)$ vanishes at the horizon, which
originates from regularity. Then we have $H(r_{0})=0$ and
\begin{equation}
h_{0}=\frac{\alpha\mu_{0}Q^{2}}{10r_{0}^{4}}-\frac{3}{2}\alpha\mu_{0}.
\end{equation}
\item The third requirement is that the charge density remains invariant. From the modified Maxwell
equation it can be seen that the charge density is proportional to $(\ast X)_{xy}$, where $X$
satisfies $\nabla_{a}X^{ab}=0$. Thus we can see that the additional terms in the modified Maxwell equation,
which come from the higher-order corrections, should not contribute
\begin{equation}
\lim_{r\rightarrow\infty}[\sqrt{-g}\alpha L^{2}(2{R_{rt}}^{rt}F_{rt}-2R^{r}_{r}F_{rt}+2R^{t}_{t}F_{tr}+RF_{rt})]=0,~~~\Rightarrow~~~h_{1}=0.
\end{equation}
\end{itemize}
Now we have fixed all the integration constants and have arrived at the complete solution.
One additional point is that, in order for the perturbative solution to be valid, the leading-order
correction to $f_{0}$, denoted as $Y(r)=f_{0}(r)F(r)$, must be much smaller than $f_{0}$.  As one approaches
the horizon $r\rightarrow r_{0}$, it can be seen that
$$F(r)|_{r\rightarrow r_{0}}=\frac{Y(r)}{f_{0}(r)}|_{r\rightarrow r_{0}}
=\frac{2\alpha Q^{2}(3Q^{2}-r_{0}^{4})}{r_{0}^{4}(Q^{2}-3r^{4}_{0})}.$$
Therefore the right hand side of the above expression must be much smaller than 1, which 
holds when $\alpha\ll1$ and $Q^{2}-3r^{4}_{0}\neq0$.  However, in the extremal limit
$Q^{2}=3r^{4}_{0}$, the denominator vanishes, which means that the 
perturbation scheme breaks down at $r=r_{0}$.

Since we have already known the perturbative solution, it is straightforward to discuss its thermodynamics.
Thermodynamics of charged black holes was extensively investigated in e.g.~\cite{Chamblin:1999tk, Chamblin:1999hg, Peca:1998cs, Anninos:2008sj}. First the temperature is given by
\begin{eqnarray}
T&=&\frac{1}{4\pi}\frac{1}{\sqrt{-g_{tt}g_{rr}}}\frac{d}{dr}g_{tt}|_{r=r_{0}}\nonumber\\
&=&\frac{1}{4\pi L^{2}}[(\frac{3M}{r_{0}^{2}}-\frac{4Q^{2}}{r_{0}^{3}})+\frac{2\alpha Q^{2}}{r_{0}^{3}}(1-\frac{3Q^{2}}{r_{0}^{4}})].
\end{eqnarray}
Next the chemical potential is still determined by the value of the gauge potential at the infinity
\begin{equation}
\mu=A_{t}(r\rightarrow\infty)=\mu_{0}-\alpha\mu_{0}(\frac{3}{2}-\frac{Q^{2}}{10r^{4}_{0}}).
\end{equation}
The other thermodynamic quantities can be evaluated in a standard way. For our particular case, the simplest
method is the so-called background subtraction method. That is, we choose some reference background and subtract
its contribution from the on-shell Euclidean action, so that the resulting Euclidean action becomes finite.
We work in the grand canonical ensemble, which means that the chemical potential is fixed. In this case we can
simply choose pure $AdS_{4}$ as the reference background. Then the remaining thermodynamic quantities can
be easily obtained. The energy density is
\begin{equation}
\epsilon=\left(\frac{\partial I_{\rm E}}{\partial\beta}\right)_{\mu}-\frac{\mu}{\beta}\left(\frac{\partial I_{\rm E}}{\partial\mu}\right)_{\beta}
=\frac{M}{\kappa^{2}L^{4}}-\alpha\frac{29Q^{4}+5Q^{2}r_{0}^{4}}{5\kappa^{2}L^{4}r_{0}^{5}}.
\end{equation}
The charge density is
\begin{equation}
\rho=-\frac{1}{\beta}\left(\frac{\partial I_{\rm E}}{\partial\mu}\right)_{\beta}=
\frac{2Q}{g_{F}\kappa^{2}L^{2}}+\frac{2\alpha(-29Q^{5}+Q^{3}r_{0}^{4})}
{5g_{F}\kappa^{2}L^{2}r_{0}^{4}(Q^{2}+3r_{0}^{4})},
\end{equation}
and the entropy density is given by
\begin{equation}
\label{entroden}
s=\beta\left(\frac{\partial I}{\partial\beta}\right)_{\mu}-I_{\rm E}=
\frac{2\pi r_{0}^{2}}{\kappa^{2}L^{2}}+\frac{2\pi\alpha Q^{2}(33Q^{2}-5r_{0}^{4})}
{5\kappa^{2}L^{2}r_{0}^{2}(Q^{2}+3r_{0}^{4})},
\end{equation}
To describe the stability of the perturbative solution, one can consider the following two quantities. One is the
specific heat $C_{\mu}$, which is defined as
$$C_{\mu}=T\left(\frac{\partial s}{\partial T}\right)_{\mu}.$$
If the specific heat is positive, the black hole is locally stable to thermal fluctuations. For our particular
case, the specific heat turns out to be
\begin{equation}
C_{\mu}=\frac{4\pi r_{0}^{2}(3r_{0}^{4}-Q^{2})}
{\kappa^{2}L^{2}(Q^{2}+3r_{0}^{4})}+\alpha\frac{4\pi Q^{2}(Q^{6}-527Q^{4}r_{0}^{4}+567Q^{2}r_{0}^{8}+135r_{0}^{12})}
{5\kappa^{2}L^{2}r_{0}^{2}(Q^{2}+3r_{0}^{4})^{3}}.
\end{equation}
The other is the electrical permittivity
$$\epsilon_{T}=\left(\frac{\partial \rho}{\partial\mu}\right)_{T},$$
whose positivity signifies the stability of the system under electrical fluctuations. Here we have
\begin{equation}
\epsilon_{T}=\frac{6r_{0}(Q^{2}+r_{0}^{4})}{g^{2}_{F}\kappa^{2}(Q^{2}+3r_{0}^{4})}
+\alpha\frac{6(-39Q^{6}+247Q^{4}r_{0}^{4}+11Q^{2}r_{0}^{8}+45r_{0}^{12})}
{10g^{2}_{F}\kappa^{2}r_{0}^{3}(Q^{2}+3r_{0}^{4})^{2}}.
\end{equation}

Let us focus on the leading order results first. According to the third law of thermodynamics, the
temperature of the black hole cannot be negative, which leads to $Q^{2}\leqslant3r_{0}^{4}$.
Therefore it can be easily seen that at the leading order, $C_{\mu}\geqslant0, \epsilon_{T}>0$, which
means that the black hole is locally stable under both thermal and electrical fluctuations.
Once higher-order corrections are taken into account, it is better to explicitly plot the curves of $C_{\mu}$
and $\epsilon_{T}$. Two specific examples are shown in the following, where we have fixed the ratio
$Q/r_{0}$ to different values since we are working in the grand canonical ensemble. Our figures show that
both $C_{\mu}$ and $\epsilon_{T}$ are definitely positive, which means that the black hole is still
thermodynamically and electrically stable at least in our specific examples.
\begin{figure}
\begin{center}
\vspace{3cm} \hspace{-0.5cm}
\includegraphics[angle=0,width=0.4\textwidth]{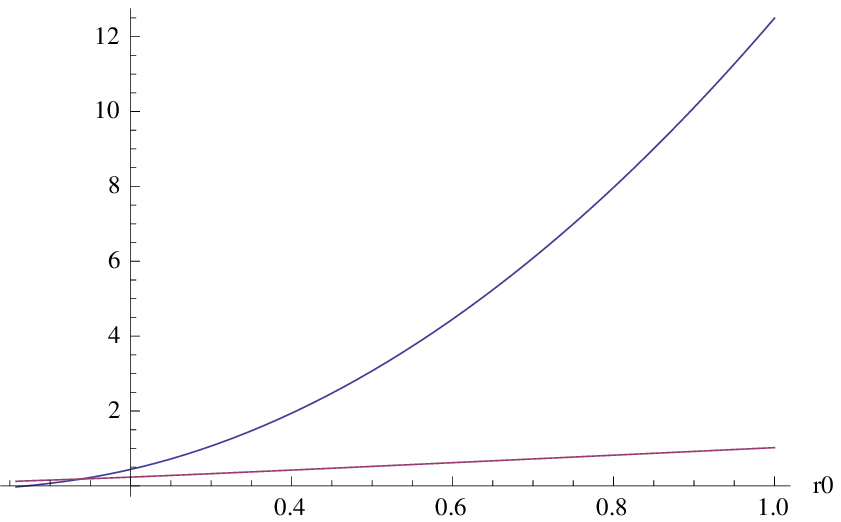}
\hspace{0.5cm}
\includegraphics[angle=0,width=0.4\textwidth]{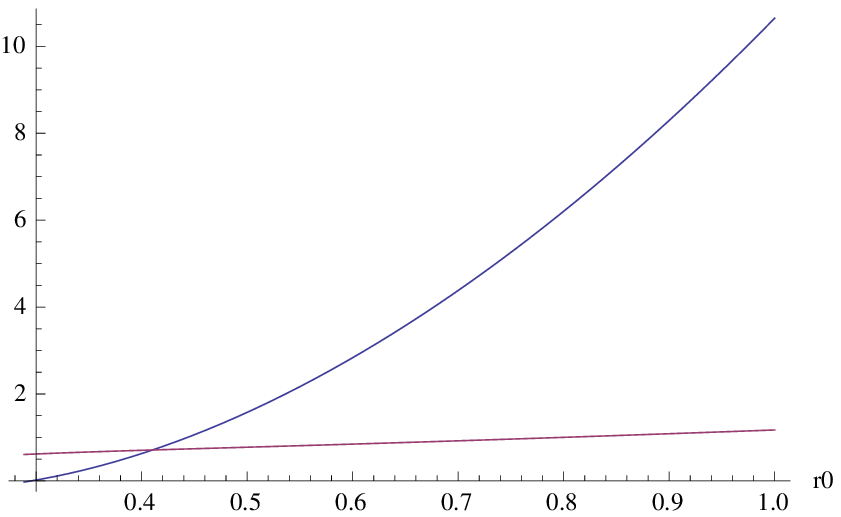}
\caption{\small The specific heat (blue line) and electrical permittivity (purple line).
Left: $\alpha=0.01, Q=0.1r_{0}$. Right: $\alpha=0.01, Q=0.5r_{0}$. The other parameters
are set to be $g_{F}=\kappa=L=1$ for simplicity.}
\end{center}
\end{figure}
\section{DC Conductivity}
The DC conductivity plays an important role in characterizing charge transport properties of
a certain system. In the hydrodynamic limit, it can be determined in terms of
the retarded current-current correlation function
\begin{equation}
\label{4eq1}
\sigma_{\rm DC}=-\lim_{\omega\rightarrow0}\frac{1}{\omega}{\rm Im}G^{R}_{xx}(\omega,\vec{k}=0),
\end{equation}
where
\begin{equation}
G^{R}_{xx}(\omega,\vec{k}=0)=-i\int dtd\vec{x}e^{i\omega t}\theta(t)\langle[J_{x}(x),J_{x}(0)]\rangle.
\end{equation}
Here $J_{x}$ denotes the CFT current dual to the bulk gauge field $A_{x}$.

It is simpler to evaluate conductivities in neutral backgrounds, as the fluctuations of the gauge field
and the fluctuations of the metric get decoupled. The conductivity in general neutral backgrounds
was investigated in~\cite{Kovtun:2008kx}, where possible bounds on the conductivity were also proposed.
Moreover, Weyl corrections to the conductivity in general dimensions were explored in~\cite{Ritz:2008kh}.
However, things become more involved in the presence of a nontrivial $U(1)$ background gauge field,
as the gauge field perturbation $A_{x}$ can couple to the metric perturbations $h_{xi}$. To deal with such a more complicated situation, the most convenient way is the effective action approach adopted in~\cite{Myers:2009ij}.
The strategy is to impose gauge invariance, which will result in a relation between the two sets
of perturbations. Then we make use of this relation to integrate out the $h_{xi}$ components and obtain an action
that involves only the $A_{x}$ fluctuation.

Before proceeding, let us introduce a new radial coordinate $u=r_{0}/r$. Then the metric
can be recast into the following form
\begin{equation}
ds^{2}=-\frac{r_{0}^{2}}{L^{2}u^{2}}f(u)dt^{2}+\frac{L^{2}du^{2}}{u^{2}f(u)}
+\frac{r_{0}^{2}}{L^{2}u^{2}}(dx^{2}+dy^{2}),
\end{equation}
where
\begin{eqnarray}
& &f(u)=(1-u)[F(u)+\alpha G(u)],~~~F(u)=1+u+u^{2}-\frac{Q^{2}u^{3}}{r_{0}^{4}},\nonumber\\
& &G(u)=\frac{Q^{2}u^{3}}{10r_{0}^{8}}
[5r_{0}^{4}(1+u+u^{2}+u^{3})-Q^{2}(11+11u+11u^{2}+11u^{3}+16u^{4})],
\end{eqnarray}
Next the fluctuations of metric components and the gauge field can be written as
\begin{eqnarray}
{h_{t}}^{x}&=&\int\frac{d^{3}k}{(2\pi)^{3}}t_{k}(u)e^{-i\omega t+iky},\nonumber\\
{h_{u}}^{x}&=&\int\frac{d^{3}k}{(2\pi)^{3}}h_{k}(u)e^{-i\omega t+iky},\nonumber\\
A_{x}&=&\int\frac{d^{3}k}{(2\pi)^{3}}a_{k}(u)e^{-i\omega t+iky},
\end{eqnarray}
where we have fixed the three-dimensional wave vector to be $\vec{k}=(-\omega,0,k)$
due to rotation symmetry in the $xy$-plane.

We consider the leading order solution, i.e. RN-${\rm AdS}_{4}$ black hole, for illustration.
Thus we have
\begin{equation}
f(u)=f_{0}(u)=(1-u)F(u),
\end{equation}
Furthermore, according to the formula for the DC conductivity~(\ref{4eq1}), it is sufficient
to set $k=0$ in subsequent calculations. Then the Maxwell equation reads
\begin{equation}
\partial_{u}[\frac{r_{0}}{L^{2}}(f_{0}(u)a^{\prime}_{k}+t_{k}A^{\prime}_{t})]
+\frac{L^{2}\omega^{2}}{r_{0}f_{0}(u)}a_{k}=0,
\end{equation}
where prime denotes the derivative with respect to $u$. To eliminate contributions from
the metric perturbations, we impose a gauge choice ${h_{u}}^{x}=0$. Then the consistency
requirement from the Einstein equation gives
\begin{equation}
\label{tk}
t_{k}^{\prime}=-4\frac{L^{4}u^{2}}{g^{2}_{F}r^{2}_{0}}A^{\prime}_{t}a_{k}.
\end{equation}
Plugging~(\ref{tk}) back into the Maxwell equation, we arrive at
\begin{equation}
\label{max}
a^{\prime\prime}_{k}+\frac{f^{\prime}_{0}(u)}{f_{0}(u)}a^{\prime}_{k}
-\frac{4u^{2}Q^{2}}{r_{0}^{4}f_{0}(u)}a_{k}+\frac{L^{4}\omega^{2}}{r_{0}^{2}f_{0}(u)^{2}}a_{k}=0,
\end{equation}
which is an equation purely for $a_{k}(u)$.

The quadratic action for the gauge field fluctuations can always be written in the following form
\begin{equation}
I^{(2)}_{a}=\frac{1}{2\kappa^{2}}\int\frac{d^{3}k}{(2\pi)^{3}}du(N(u)a^{\prime}_{k}a^{\prime}_{-k}
+M(u)a_{k}a_{-k}),
\end{equation}
where
\begin{equation}
N(u)=-\frac{r_{0}}{g^{2}_{F}}f_{0}(u),~~~
M(u)=\frac{L^{4}\omega^{2}}{r_{0}g^{2}_{F}f_{0}(u)}-\frac{4L^{4}u^{2}}{r_{0}g_{F}^{4}}A^{\prime2}_{t}
\end{equation}
can be read off by making use of~(\ref{max}). The general form of the equation of motion can be
expressed as
\begin{equation}
\partial_{u}j_{k}(u)=\frac{1}{\kappa^{2}}M(u)a_{k}(u),
\end{equation}
where
\begin{equation}
j_{k}(u)\equiv\frac{\delta I^{(2)}_{a}}{\delta a^{\prime}_{-k}}=\frac{1}{\kappa^{2}}N(u)a^{\prime}_{k}(u)
\end{equation}
denotes the canonical momentum.

To proceed, one can impose the following regularity condition at the horizon
$u=u_{0}$, which originates from the black hole membrane paradigm~\cite{Iqbal:2008by},
\begin{equation}
\label{mem}
j_{k}(u_{0})=-i\omega\lim_{u\rightarrow u_{0}}\frac{N(u)}{\kappa^{2}}\sqrt{\frac{g_{uu}}{-g_{tt}}}
a_{k}(u)+\mathcal{O}(\omega^{2}).
\end{equation}
Note that here we are expanding in the low-frequency limit. Next one can easily evaluate
the flux factor from the on-shell boundary action~\cite{Son:2002sd}
\begin{equation}
2\mathcal{F}_{k}=j_{k}(u)a_{-k}(u).
\end{equation}
Then according to~(\ref{4eq1}), the DC conductivity is given by
\begin{equation}
\sigma_{\rm DC}=\lim_{u,\omega\rightarrow0}\frac{1}{\omega}{\rm Im}\left[\frac{2\mathcal{F}_{k}}{a_{k}(u)a_{-k}(u)}\right]_{k=0}
=\lim_{u,\omega\rightarrow0}{\rm Im}\left[\frac{j_{k}(u)a_{-k}(u)}{\omega a_{k}(u)a_{-k}(u)}\right]_{k=0},
\end{equation}

One crucial point is that the numerator in the previous formula is a conserved quantity,
\begin{equation}
\frac{d}{du}{\rm Im}[j_{k}(u)a_{-k}(u)]={\rm Im}(f_{1}(u)a_{k}(u)a_{-k}(u)+f_{2}(u)j_{k}(u)j_{-k}(u))=0,
\end{equation}
as the resulting two terms are real. Since it does not evolve along the radial direction,
it can be evaluated at any point in the spacetime. We calculate this quantity at the horizon,
where the regularity condition~(\ref{mem}) should be imposed. Finally the DC conductivity
can be written as
\begin{equation}
\sigma_{\rm DC}=\frac{1}{\kappa^{2}}K^{2}_{A}(u_{0})\frac{\mathcal{N}(u_{0})}{\mathcal{N}(0)}|_{k=0},
\end{equation}
where
\begin{equation}
K^{2}_{A}(u)=-N(u)\sqrt{\frac{g_{uu}}{-g_{tt}}},~~~\mathcal{N}(u)=a_{k}(u)a_{-k}(u).
\end{equation}

Notice that here $\mathcal{N}(u)$ is also real and so is independent of $\omega$ up to
$\mathcal{O}(\omega^{2})$, which means that up to this order $\mathcal{N}(u)$ is regular
at the horizon. Then in order to evaluate $\mathcal{N}$, it is sufficient to solve for $a_{k}(u)$
by setting $\omega=0$ in the corresponding equation of motion. The calculations can be considerably
simplified and the solution can be easily obtained
\begin{equation}
a_{k}(u)=a_{0}(1-\frac{4Q^{2}}{3(r_{0}^{4}+Q^{2})}u).
\end{equation}
Finally the DC conductivity is given by
\begin{equation}
\sigma_{\rm DC}=\frac{L^{2}}{\kappa^{2}g^{2}_{F}}\frac{(3r^{4}_{0}-Q^{2})^{2}}{9(r^{4}_{0}+Q^{2})^{2}},
\end{equation}
which agrees with the previous result, see e.g.~\cite{Ge:2010yc}.

Our next task is to evaluate the effects of higher-order corrections. The steps are indeed
more or less the same. To be concrete, we impose the gauge choice ${h_{u}}^{x}=0$ and eliminate contributions
from metric perturbations, which leads to an equation purely for the gauge field fluctuation.
Then the DC conductivity can be obtained in the same way. However, the key problem is that
the equation becomes more complicated and it is very difficult to find any analytic solution.
Therefore we just keep the solution up to the first order of $\alpha$ and $Q^{2}$, which reads
\begin{equation}
a_{k}(u)=a_{0}+a_{1}u+\alpha[\frac{2}{3}a_{0}u(2+u)
+\frac{1}{2}a_{1}u(u-2)],~~~a_{1}=-\frac{4a_{0}Q^{2}}{3(r_{0}^{4}+Q^{2})}.
\end{equation}
Thus the DC conductivity at the first order is given by
\begin{equation}
\sigma_{\rm DC}=\frac{L^{2}}{\kappa^{2}g^{2}_{F}}\left[\frac{(3r^{4}_{0}-Q^{2})^{2}}{9(r^{4}_{0}+Q^{2})^{2}}
+\alpha\left(4+\frac{4Q^{2}}{3(r_{0}^{4}+Q^{2})}\right)\left(1-\frac{4Q^{2}}{3(r_{0}^{4}+Q^{2})}\right)\right],
\end{equation}
When we consider the limit $Q=0$, the DC conductivity becomes
\begin{equation}
\sigma_{\rm DC}=\frac{L^{2}}{\kappa^{2}g^{2}_{F}}(1+4\alpha).
\end{equation}
It can be seen that the result obtained in~\cite{Myers:2010pk} can be reproduced even at higher-order.
\section{Shear viscosity, thermal conductivity and relevant ratios}
In this section we discuss two other interesting quantities, the shear viscosity $\eta$ and
the thermal conductivity $\kappa_{T}$, as well as the relevant ratios, $\eta/s$ and
$\kappa_{T}\mu^{2}/(\eta T)$. At first sight, the shear viscosity in three-dimensional
field theory spacetime does not seem to be as interesting as the four-dimensional counterpart.
However, it would be worth investigating as this quantity is directly related to the ratio
$\kappa_{T}\mu^{2}/(\eta T)$.

The shear viscosity is related to the low frequency and zero momentum limit of the retarded
Green's function of the stress tensor in the dual CFT via Kubo's formula
\begin{equation}
\eta=-\lim_{\omega\rightarrow0}\frac{1}{\omega}{\rm Im}G^{R}_{xy,xy}(\omega,\vec{k}=0).
\end{equation}
Here the retarded Green's function is defined as
\begin{equation}
G^{R}_{xy,xy}(\omega,\vec{k}=0)=-i\int dtd\vec{x}e^{i\omega t}\theta(t)\langle[T_{xy}(x),T_{xy}(0)]\rangle,
\end{equation}
By now there have been several methods for evaluating the shear viscosity in the presence of higher-derivative corrections. An incomplete list of references is shown in~\cite{incomplete} but here we still
adopt the effective action approach in~\cite{Myers:2009ij}. The ansatz for the metric perturbation
is
\begin{equation}
{h_{x}}^{y}(t,u)=\int\frac{d^{3}k}{(2\pi)^{3}}\phi(u)e^{-i\omega t},
\end{equation}
where we have set $k=0$.

The strategy here is similar to that in the calculations of the DC conductivity.
We also expand the action to quadratic order in $\phi$ and read off the shear viscosity
via the membrane paradigm. For a general theory of gravity, where higher-order corrections
may exist, the quadratic effective action for $\phi$ can always be expressed in the following
form
\begin{eqnarray}
I^{(2)}_{\phi}&=&\frac{1}{2\kappa^{2}}\int\frac{d^{3}k}{(2\pi)^{3}}du
[A(u)\phi^{\prime\prime}\phi+B(u)\phi^{\prime}\phi^{\prime}+C(u)\phi^{\prime}\phi\nonumber\\
& &+D(u)\phi\phi+E(u)\phi^{\prime\prime}\phi^{\prime\prime}+F\phi^{\prime\prime}\phi^{\prime}+K_{\rm GH}],
\end{eqnarray}
where $K_{\rm GH}$ denotes contributions from the Gibbons-Hawking terms.
Subsequently, by making use of the equation of motion and integrating by parts, we arrive
at the following `modified' effective action
\begin{equation}
\tilde{I}^{(2)}_{\phi}=\frac{1}{2\kappa^{2}}\int\frac{d^{3}k}{(2\pi)^{3}}du
[(B-A-\frac{F^{\prime}}{2})\phi^{\prime}\phi^{\prime}+E(u)\phi^{\prime\prime}\phi^{\prime\prime}
+(D-\frac{(C-A^{\prime})^{\prime}}{2})\phi\phi]+\tilde{K}_{\rm GH},
\end{equation}
where the Gibbons-Hawking term is also modified.

Given the `modified' effective action, it is convenient to define the canonical momentum
for the effective scalar
\begin{equation}
\Pi(u)\equiv\frac{\delta\tilde{I}^{(2)}_{\phi}}{\delta\phi^{\prime}}
=\frac{1}{\kappa^{2}}[(B-A-\frac{F^{\prime}}{2})-(E(u)\phi^{\prime\prime})^{\prime}].
\end{equation}
Notice that here the `modified' Gibbons-Hawking term does not contribute
to the canonical momentum. As demonstrated in~\cite{Iqbal:2008by}, in the low
frequency limit the shear viscosity is determined by the canonical momentum as follows
\begin{equation}
\eta=\lim_{u,\omega\rightarrow0}\frac{\Pi(u)}{i\omega\phi(u)}.
\end{equation}
Furthermore, the equation of motion for the effective scalar is given by
\begin{equation}
\partial_{u}\Pi(u)=M(u)\phi(u),~~~M(u)\equiv\frac{1}{\kappa^{2}}(D-(C-A^{\prime})^{\prime}/2).
\end{equation}
It can be seen that $M(u)\sim\mathcal{O}(\omega^{2})$, therefore
in the low frequency limit $\partial_{u}\Pi(u)=0$. Hence $\Pi(u)$ is also a conserved quantity and it
can be evaluated at any radial position. Here we evaluate $\Pi(u)$ at the horizon once again
and impose the regularity condition for the effective scalar $\phi$. Finally the shear viscosity
is given by
\begin{equation}
\eta=\frac{1}{\kappa^{2}}(K^{2}_{\phi}(u_{0})+K^{4}_{\phi}(u_{0})),
\end{equation}
where
\begin{equation}
K^{(2)}_{\phi}(u)=\sqrt{\frac{g_{uu}}{-g_{tt}}}(A-B+\frac{F^{\prime}}{2}),~~~
K^{(4)}_{\phi}(u)=\left[E(u)\left(\sqrt{\frac{g_{uu}}{-g_{tt}}}\right)^{\prime}\right]^{\prime},
\end{equation}

The above calculations are rather general. However, for our particular background,
the non-vanishing functions in $\tilde{I}^{(2)}_{\phi}$ are given as follows
\begin{eqnarray}
& &A(u)=\frac{2r_{0}^{4}f_(u)}{L^{4}u^{2}},~~~B(u)=\frac{3r^{3}_{0}f(u)}{2L^{4}u^{2}}
-\frac{\alpha u^{2}Q^{2}f(u)}{L^{4}r_{0}},\nonumber\\
& &C(u)=-\frac{6r_{0}^{3}f(u)}{L^{4}u^{3}}+\frac{2r_{0}^{3}f(u)^{\prime}}{L^{4}u^{2}}
-\frac{4\alpha uQ^{2}f(u)}{L^{4}r_{0}}.
\end{eqnarray}
Therefore
\begin{equation}
K^{2}_{\phi}(u)=\frac{r_{0}^{2}}{2u^{2}L^{2}}+\frac{\alpha u^{2}Q^{2}}{L^{2}r_{0}^{2}},
~~~K^{4}_{\phi}(u)=0,
\end{equation}
which lead to
\begin{equation}
\eta=\frac{1}{\kappa^{2}}(\frac{r_{0}^{2}}{2L^{2}}+\frac{\alpha Q^{2}}{L^{2}r_{0}^{2}}).
\end{equation}
Recalling the entropy density given in~(\ref{entroden}), we arrive at
\begin{equation}
\frac{\eta}{s}=\frac{1}{4\pi}\left[1+\frac{\alpha Q^{2}(35r_{0}^{4}-23Q^{2})}{5r_{0}^{4}(Q^{2}+3r_{0}^{4})}\right].
\end{equation}
From this result, it can be seen that the bound $1/4\pi$ might be violated, depending
on the sign of the gauge coupling parameter $\alpha$. Furthermore, when the charge
parameter $Q$ goes to zero, we can recover $1/4\pi$.

Another interesting quantity is the thermal conductivity, which describes the response
of the heat flow to temperature gradients, ${T^{t}}_{i}=-\kappa_{T}\partial_{i}T$.
As stated in~\cite{Son:2006em}, the thermal conductivity can be expressed as follows
 \begin{equation}
\kappa_{T}=(\frac{s}{\rho}+\frac{\mu}{T})^{2}T\sigma_{\rm DC}.
\end{equation}
Furthermore, thermal conductivity was systematically investigated in~\cite{Jain:2009pw}
in the context of holography. Since we have already obtained all the quantities in the
above expression, it is straightforward to calculate the thermal conductivity.
Moreover, we can evaluate another ratio $\kappa_{T}\mu^{2}/(\eta T)$.
It was found in~\cite{Son:2006em} that $\kappa_{T}\mu^{2}/(\eta T)=8\pi^2$ at the leading
order, which was an analogue of the Wiedemann-Franz law. Such a value can also be
modified by higher-order corrections~\cite{Myers:2009ij}.
For our case, the ratio is given by
\begin{equation}
\frac{\kappa_{T}\mu^{2}}{\eta T}=
2\pi^{2}g^{2}_{F}+\alpha\pi^{2}g_{F}^{2}(\frac{154Q^{2}}{5r_{0}^{4}}-2),
\end{equation}
where the differences in the leading order result are due to different normalizations
in the gauge kinetic term. One can see that this ratio can also be modified by higher-order
corrections.
\section{Summary and discussion}
Charge transport properties are of particular interest in condensed matter physics,
thus it would be desirable to investigate them in the dual gravity side in the spirit
of the AdS/CMT correspondence. In~\cite{Myers:2010pk}, the authors studied
charge transport properties in Schwarzschild-${\rm AdS}_{4}$ background in the
presence of a particular form of higher-order corrections~(\ref{equ2}). Furthermore,
it was observed that the $RF^{2}$ higher-order corrections in~(\ref{equ4}), which originated
from Kaluza-Klein reduction of five-dimensional Gauss-Bonnet gravity, were equivalent
to~(\ref{equ2}) in neutral backgrounds, which led to the same charge transport properties.
However, when the background is at finite charge density, the two above mentioned higher-order corrections are no longer equivalent. Then in order to understand the effects of
higher-order corrections to charge transport properties at finite density,
we have to perform case-by-case analysis.

In this paper we study several holographic aspects of charged black holes under
$RF^{2}$ corrections shown in~(\ref{equ4}), which can be seen as a primary step toward
better understanding of charge transport properties at finite density. Due to the presence
of the background gauge field, it is necessary to work out the full solution in the beginning,
at least perturbatively. We obtain the perturbative solution and discuss its thermodynamics
and local stability. Next we calculate the DC conductivity using the effective action
approach proposed in~\cite{Myers:2009ij}. We find that at the leading order the result agrees with
that obtained in other literatures and we also obtain corrections to the DC conductivity.
In particular, if we take the $Q=0$ limit, our result can reproduce that obtained
in~\cite{Myers:2010pk}. Subsequently we compute two additional
hydrodynamic quantities, the shear viscosity $\eta$ and the thermal conductivity $\kappa_{T}$.
It can be seen that the higher-order corrections can affect both $\eta/s$ and
$\kappa_{T}\mu^{2}/(\eta T)$. However, the conjectured bound $\eta/s=1/4\pi$ can be recovered
when $Q=0$.

One topic that we do not investigate in detail in the main text is the conductivity at extremality.
At zero temperature the near horizon geometry is $AdS_{2}\times R^{2}$ and the low-frequency
behavior of the retarded Green's function is determined  by an emergent IR $CFT_{1}$~\cite{Faulkner:2009wj}.
Furthermore, Wald-like formulae for the shear viscosity and the conductivity at extremality in general
higher-derivative gravity theories were proposed in~\cite{Paulos:2009yk}. It was found in~\cite{Paulos:2009yk}
that the real part of the conductivity had a universal behavior ${\rm Re}\sigma\propto\omega^{2}$ under various
forms of higher-order corrections, including the $RF^{2}$ terms we consider in this paper. This universality
supported the argument that the low-frequency behavior of the retarded Green's functions is determined by the IR CFT.
Transport coefficients in extremal black hole backgrounds were also investigated in~\cite{extremal}.

To have a more comprehensive understanding on charge transport properties, it is necessary
to work out the full retarded Green's functions. However, once again due to the nontrivial
background $U(1)$ gauge field, it is more difficult to compute the corresponding correlation
functions even at the leading order, see e.g.~\cite{Ge:2010yc}. Then one may resort to perturbative
methods and numerical techniques to obtain such correlation functions. Moreover, once we know the full correlation functions at the first order, we can discuss the effects of higher-order corrections to e.g. holographic optics~\cite{Amariti:2010jw, Amariti:2010hw}. We expect to report progress in the above mentioned directions in the future.

\bigskip \goodbreak \centerline{\bf Acknowledgments}
\noindent DWP would like to thank Institute of Theoretical Physics,
Chinese Academy of Sciences for hospitality, where most of the work
was done, as well as Song He and Jose' Sande Lemos for comments.
This work was supported in part by the National Natural
Science Foundation of China (No. 10821504, No. 10975168 and
No.11035008), and in part by the Ministry of Science and Technology
of China under Grant No. 2010CB833004. DWP acknowledges an FCT
(Portuguese Science Foundation) grant. This work was also funded by
FCT through projects CERN/FP/109276/2009 and PTDC/FIS/098962/2008.




\begin{thebibliography}{99}
\addcontentsline{toc}{section}{References}


\bibitem{Maldacena:1997re}
  J.~M.~Maldacena,
  ``The large N limit of superconformal field theories and supergravity,''
  Adv.\ Theor.\ Math.\ Phys.\  {\bf 2}, 231 (1998)
  [Int.\ J.\ Theor.\ Phys.\  {\bf 38}, 1113 (1999)]
  [arXiv:hep-th/9711200].\\
  S.~S.~Gubser, I.~R.~Klebanov and A.~M.~Polyakov,
  ``Gauge theory correlators from non-critical string theory,''
  Phys.\ Lett.\  B {\bf 428}, 105 (1998)
  [arXiv:hep-th/9802109].\\
  E.~Witten,
  ``Anti-de Sitter space and holography,''
  Adv.\ Theor.\ Math.\ Phys.\  {\bf 2}, 253 (1998)
  [arXiv:hep-th/9802150].

\bibitem{Aharony:1999ti}
  O.~Aharony, S.~S.~Gubser, J.~M.~Maldacena, H.~Ooguri and Y.~Oz,
  ``Large N field theories, string theory and gravity,''
  Phys.\ Rept.\  {\bf 323}, 183 (2000)
  [arXiv:hep-th/9905111].

\bibitem{Hartnoll:2009sz}
  S.~A.~Hartnoll,
  ``Lectures on holographic methods for condensed matter physics,''\\
  Class.\ Quant.\ Grav.\  {\bf 26}, 224002 (2009)
  [arXiv:0903.3246 [hep-th]].\\
  C.~P.~Herzog,
  ``Lectures on Holographic Superfluidity and Superconductivity,''\\
  J.\ Phys.\ A  {\bf 42}, 343001 (2009)
  [arXiv:0904.1975 [hep-th]].\\
   J.~McGreevy,
  ``Holographic duality with a view toward many-body physics,''\\
  arXiv:0909.0518 [hep-th].\\
  G.~T.~Horowitz,
  ``Introduction to Holographic Superconductors,''\\
  arXiv:1002.1722 [hep-th].\\
  S.~Sachdev,
  ``Condensed matter and AdS/CFT,''
  arXiv:1002.2947 [hep-th].
\bibitem{Lee:2008xf}
  S.~S.~Lee,
  ``A Non-Fermi Liquid from a Charged Black Hole: A Critical Fermi Ball,''
  Phys.\ Rev.\  D {\bf 79}, 086006 (2009)
  [arXiv:0809.3402 [hep-th]].

\bibitem{Liu:2009dm}
H.~Liu, J.~McGreevy and D.~Vegh, ``Non-Fermi liquids from holography,''
arXiv:0903.2477 [hep-th].

\bibitem{Cubrovic:2009ye}
  M.~Cubrovic, J.~Zaanen and K.~Schalm,
  ``String Theory, Quantum Phase Transitions and the Emergent Fermi-Liquid,''
  Science {\bf 325}, 439 (2009)
  [arXiv:0904.1993 [hep-th]].

\bibitem{Faulkner:2009wj}
  T.~Faulkner, H.~Liu, J.~McGreevy and D.~Vegh,
  ``Emergent quantum criticality, Fermi surfaces, and AdS2,''
  arXiv:0907.2694 [hep-th].
\bibitem{Gubser:2008px}
  S.~S.~Gubser,
  ``Breaking an Abelian gauge symmetry near a black hole horizon,''
  Phys.\ Rev.\  D {\bf 78}, 065034 (2008)
  [arXiv:0801.2977 [hep-th]].
\bibitem{Hartnoll:2008vx}
  S.~A.~Hartnoll, C.~P.~Herzog and G.~T.~Horowitz,
  ``Building a Holographic Superconductor,''
  Phys.\ Rev.\ Lett.\  {\bf 101}, 031601 (2008)
  [arXiv:0803.3295 [hep-th]].
\bibitem{Hartnoll:2008kx}
  S.~A.~Hartnoll, C.~P.~Herzog and G.~T.~Horowitz,
  ``Holographic Superconductors,''
  JHEP {\bf 0812}, 015 (2008)
  [arXiv:0810.1563 [hep-th]].

\bibitem{Herzog:2007ij}
  C.~P.~Herzog, P.~Kovtun, S.~Sachdev and D.~T.~Son,
  ``Quantum critical transport, duality, and M-theory,''
  Phys.\ Rev.\  D {\bf 75}, 085020 (2007)
  [arXiv:hep-th/0701036].

\bibitem{Myers:2010pk}
  R.~C.~Myers, S.~Sachdev and A.~Singh,
  ``Holographic Quantum Critical Transport without Self-Duality,''
  arXiv:1010.0443 [hep-th].

\bibitem{Myers:2009ij}
  R.~C.~Myers, M.~F.~Paulos and A.~Sinha,
  ``Holographic Hydrodynamics with a Chemical Potential,''
  JHEP {\bf 0906}, 006 (2009)
  [arXiv:0903.2834 [hep-th]].
\bibitem{Cremonini:2008tw}
  S.~Cremonini, K.~Hanaki, J.~T.~Liu and P.~Szepietowski,
  ``Black holes in five-dimensional gauged supergravity with higher
  derivatives,''
  JHEP {\bf 0912}, 045 (2009)
  [arXiv:0812.3572 [hep-th]].
\bibitem{Chamblin:1999tk}
  A.~Chamblin, R.~Emparan, C.~V.~Johnson, R.~C.~Myers,
  ``Charged AdS black holes and catastrophic holography,''
  Phys.\ Rev.\  {\bf D60}, 064018 (1999).
  [hep-th/9902170].
\bibitem{Chamblin:1999hg}
  A.~Chamblin, R.~Emparan, C.~V.~Johnson, R.~C.~Myers,
  ``Holography, thermodynamics and fluctuations of charged AdS black holes,''
  Phys.\ Rev.\  {\bf D60}, 104026 (1999).
  [hep-th/9904197].
\bibitem{Peca:1998cs}
  C.~S.~Peca, J.~Lemos, P.S.,
  ``Thermodynamics of Reissner-Nordstrom anti-de Sitter black holes in the grand canonical ensemble,''
  Phys.\ Rev.\  {\bf D59}, 124007 (1999).
  [gr-qc/9805004].
\bibitem{Anninos:2008sj}
  D.~Anninos, G.~Pastras,
  ``Thermodynamics of the Maxwell-Gauss-Bonnet anti-de Sitter Black Hole with Higher Derivative Gauge Corrections,''
  JHEP {\bf 0907}, 030 (2009).
  [arXiv:0807.3478 [hep-th]].
\bibitem{Kovtun:2008kx}
  P.~Kovtun, A.~Ritz,
  ``Universal conductivity and central charges,''
  Phys.\ Rev.\  {\bf D78}, 066009 (2008).
\bibitem{Ritz:2008kh}
  A.~Ritz, J.~Ward,
  ``Weyl corrections to holographic conductivity,''
  Phys.\ Rev.\  {\bf D79}, 066003 (2009).
  [arXiv:0811.4195 [hep-th]].
\bibitem{Iqbal:2008by}
  N.~Iqbal and H.~Liu,
  ``Universality of the hydrodynamic limit in AdS/CFT and the membrane
  paradigm,''
  Phys.\ Rev.\  D {\bf 79}, 025023 (2009)
  [arXiv:0809.3808 [hep-th]].
\bibitem{Son:2002sd}
  D.~T.~Son, A.~O.~Starinets,
  ``Minkowski space correlators in AdS / CFT correspondence: Recipe and applications,''
  JHEP {\bf 0209}, 042 (2002).
  [hep-th/0205051].
\bibitem{Ge:2010yc}
  X.~H.~Ge, K.~Jo and S.~J.~Sin,
  ``Hydrodynamics of RN AdS$_4$ black hole and Holographic Optics,''
  JHEP {\bf 1103}, 104 (2011)
  [arXiv:1012.2515 [hep-th]].
\bibitem{incomplete}
  Y.~Kats, P.~Petrov,
  ``Effect of curvature squared corrections in AdS on the viscosity of the dual gauge theory,''
  JHEP {\bf 0901}, 044 (2009).
  [arXiv:0712.0743 [hep-th]].\\
  M.~Brigante, H.~Liu, R.~C.~Myers, S.~Shenker, S.~Yaida,
  ``Viscosity Bound Violation in Higher Derivative Gravity,''
  Phys.\ Rev.\  {\bf D77}, 126006 (2008).\\
  M.~Brigante, H.~Liu, R.~C.~Myers, S.~Shenker, S.~Yaida,
  ``The Viscosity Bound and Causality Violation,''
  Phys.\ Rev.\ Lett.\  {\bf 100}, 191601 (2008).
  [arXiv:0802.3318 [hep-th]].\\
  X.~H.~Ge, Y.~Matsuo, F.~W.~Shu, S.~J.~Sin, T.~Tsukioka,
  ``Viscosity Bound, Causality Violation and Instability with Stringy Correction and Charge,''
  JHEP {\bf 0810}, 009 (2008).
  [arXiv:0808.2354 [hep-th]].\\
  R.~Brustein, A.~J.~M.~Medved,
  ``The Ratio of shear viscosity to entropy density in generalized theories of gravity,''
  Phys.\ Rev.\  {\bf D79}, 021901 (2009).
  [arXiv:0808.3498 [hep-th]].\\
  R.~G.~Cai, Z.~Y.~Nie, Y.~W.~Sun,
  ``Shear Viscosity from Effective Couplings of Gravitons,''
  Phys.\ Rev.\  {\bf D78}, 126007 (2008).
  [arXiv:0811.1665 [hep-th]].\\
  A.~Buchel, R.~C.~Myers, A.~Sinha,
  ``Beyond $\eta/s = 1/4 \pi$,''
  JHEP {\bf 0903}, 084 (2009).
  [arXiv:0812.2521 [hep-th]].\\
  R.~G.~Cai, Z.~Y.~Nie, N.~Ohta, Y.~W.~Sun,
  ``Shear Viscosity from Gauss-Bonnet Gravity with a Dilaton Coupling,''
  Phys.\ Rev.\  {\bf D79}, 066004 (2009).
  [arXiv:0901.1421 [hep-th]].
\bibitem{Son:2006em}
  D.~T.~Son, A.~O.~Starinets,
  ``Hydrodynamics of r-charged black holes,''
  JHEP {\bf 0603}, 052 (2006).
  [hep-th/0601157].
\bibitem{Jain:2009pw}
  S.~Jain,
  ``Holographic electrical and thermal conductivity in strongly coupled gauge theory with multiple chemical potentials,''
  JHEP {\bf 1003}, 101 (2010).
  [arXiv:0912.2228 [hep-th]].\\
  S.~Jain,
  ``Universal properties of thermal and electrical conductivity of gauge theory plasmas from holography,''
  JHEP {\bf 1006}, 023 (2010).
  [arXiv:0912.2719 [hep-th]].\\
  S.~Jain,
  ``Universal thermal and electrical conductivity from holography,''
  JHEP {\bf 1011}, 092 (2010).
  [arXiv:1008.2944 [hep-th]].

\bibitem{Paulos:2009yk}
  M.~F.~Paulos,
  ``Transport coefficients, membrane couplings and universality at extremality,''
  JHEP {\bf 1002}, 067 (2010).
  [arXiv:0910.4602 [hep-th]].
\bibitem{extremal}
  M.~Edalati, J.~I.~Jottar, R.~G.~Leigh,
  ``Transport Coefficients at Zero Temperature from Extremal Black Holes,''
  JHEP {\bf 1001}, 018 (2010).
  [arXiv:0910.0645 [hep-th]].\\
  R.~G.~Cai, Y.~Liu, Y.~W.~Sun,
  ``Transport Coefficients from Extremal Gauss-Bonnet Black Holes,''
  JHEP {\bf 1004}, 090 (2010).
  [arXiv:0910.4705 [hep-th]].\\
  S.~K.~Chakrabarti, S.~Jain, S.~Mukherji,
  ``Viscosity to entropy ratio at extremality,''
  JHEP {\bf 1001}, 068 (2010).
  [arXiv:0910.5132 [hep-th]].\\
  M.~Edalati, J.~I.~Jottar, R.~G.~Leigh,
  ``Shear Modes, Criticality and Extremal Black Holes,''
  JHEP {\bf 1004}, 075 (2010).
  [arXiv:1001.0779 [hep-th]].
  \bibitem{Amariti:2010jw}
  A.~Amariti, D.~Forcella, A.~Mariotti and G.~Policastro,
  ``Holographic Optics and Negative Refractive Index,''
  arXiv:1006.5714 [hep-th].
\bibitem{Amariti:2010hw}
  A.~Amariti, D.~Forcella, A.~Mariotti,
  ``Additional Light Waves in Hydrodynamics and Holography,''
   [arXiv:1010.1297 [hep-th]].


\end{thebibliography}
\end{document}